\documentclass[aps,showpacs]{revtex4}
\usepackage{amssymb}
\usepackage{graphicx}
\usepackage{dcolumn}
\usepackage{amsmath}
\begin{document}
\title{Non-quasiparticle states in a half-metallic ferromagnet with antiferromagnetic $s-d(f)$ interaction}
\author{V. Yu. Irkhin}
\email{Valentin.Irkhin@imp.uran.ru}
\affiliation{Institute of Metal Physics, 620990 Ekaterinburg, Russia}

\begin{abstract}
Non-quasiparticle (incoherent) states which play an important role in the electronic structure of half-metallic ferromagnets (HMF) are investigated consistently in the case of antiferromagnetic $s-d(f)$ exchange interaction. Their appropriate description in the limit of strong correlations requires a rearrangement of perturbation series in comparison with the usual Dyson equation. This consideration provides a solution of the Kondo problem in the HMF case and can be important for first-principle HMF calculations performed earlier for ferromagnetic $s-d(f)$ interaction.
\end{abstract}

\maketitle

\section{Introduction}

The s-d(f) exchange (Kondo lattice) model is widely used to describe many-electron (correlation) effects in $d-$ and $f-$metals and their compounds. This model at large $|I|$ is equivalent to the $t-J$ model which is applicable, e.g., for high-$T_c$ copper oxides.

Besides the Kondo effect in the paramagnetic state, the correlation effects are important for magnetic semiconductors and half-metallic ferromagnets (HMF).
The latter systems are widely investigated last time \cite{RMP}. In particular, the interest in HFM is connected with possible applications in spintronics, including giant magnetoresistance (GMR) owing to spin-polarized electron transport across a non-magnetic metallic layer \cite{Tak} (this phenomenon in HMF was predicted in Ref.\cite{UFN}).

In the simple one-band case, HFM are strong (saturated) ferromagnets where (as well as in degenerate ferromagnetic semiconductors) the spin splitting exceeds the Fermi level, so that one spin subband is empty in the Hartree-Fock (Stoner) picture. However, this picture is  radically changed by correlation effects which have in HFM a rather specific form, being connected with occurrence of incoherent non-quasiparticle (NQP) states \cite{RMP}.
These states yield appreciable contributions to various physical properties (spin polarization, specific heat, transport characteristics etc.) \cite{RMP,RMP1}.
They occur due to quantum effects and are the only states with one of spin projections ($\sigma = \downarrow $ for the positive $s-d(f)$ exchange parameter $I$ and $\sigma = \uparrow $ for $I<0$) near the Fermi level. The NQP states were also treated within first-principle calculations \cite{NQP,NQP1}.

In such a situation, the $s-d(f)$ exchange model provides an example of a non-trivial many-particle problem which can be investigated rather rigorously beyond perturbation theory. This is owing to that the ground ferromagnetic state is known exactly.
As for field-theoretical methods, the case of a half-metallic ferromagnet can be investigated rather strictly. The exact solution in the one-electron case (empty conduction band) reduces the problem to the Schroedinger equation (two-body electron-magnon scattering problem). At finite temperatures (in the spin-wave region) and electron concentrations, a consistent consideration can be performed by constructing expansion in the occupation numbers of electrons and magnons. To this end, there can be used methods of equations of motion for the Green's functions \cite{329}, of the generating functional \cite{329a}, and the expansion of the operator of evolution \cite{329b}. In this case, in each order there arise integral equations that describe the electron--magnon scattering (in their structure, they are similar to the integral equations of the Nagaoka type~\cite{349}).

However, the problem of NQP states in the case of negative (antiferromagnetic) $s-d(f)$ parameter is still not solved satisfactorily, the physical picture in this case being especially non-trivial since the states below the Fermi become depolarized owing to occupied NQP states.
In fact, the above-discussed expansion is somewhat different for ferromagnetic and antiferromagnetic $s-d(f)$ exchange parameters (the difference is similar to that in the  Kondo problem in the paramagnetic situation).

In the present paper we present a consistent solution of the negative $I$ problem and treat in detail this difference.
In Sect.2 we derive the expressions for the one-electron Green's functions, which are specific for the NQP situation.
In Sect.3 we present the numerical calculation  results and discuss them.
Appendix deals with the limit of strong correlations.

\section{Calculation of one-electron Green's functions}

We write down the Hamiltonian of the $s-f$ exchange model in the form
\begin{equation}
H= \sum_{\mathbf{k}\sigma
}t_{\mathbf{k}}c_{\mathbf{k}\sigma }^{\dagger }c_{\mathbf{k}\sigma
}+\sum_{\mathbf{q}}J_{\mathbf{q}}\mathbf{S}_{-\mathbf{q}}
\mathbf{S}_{\mathbf{q}}-I\sum_{i\sigma \sigma ^{\prime
}}(\mathbf{S}_i\mbox{\boldmath$\sigma $}_{\sigma \sigma ^{\prime
}})c_{i\sigma }^{\dagger }c_{i\sigma ^{\prime }}, \label{eq:G.2}
\end{equation}
where {\boldmath$\sigma $} are the Pauli matrices, $ t_{\mathbf{k}}$ is the bare band energy, $I$ is the $s-f$ exchange parameter.
We consider the one-electron retarded Zubarev Green's function.

\begin{equation}
G_{\mathbf{k}\sigma }(E)=\langle \!\langle c_{\mathbf{k}\sigma }|c_{\mathbf{k%
}\sigma }^{\dagger }\rangle \!\rangle _{E}  \label{eq:G.30}
\end{equation}%
We write down the Dyson equation for a ferromagnet at zero temperature
\begin{equation}
G_{\mathbf{k\sigma }}(E)=\left[ E-t_{\mathbf{k}\sigma }-\Sigma _{\mathbf{k}%
}^{\sigma}(E)\right] ^{-1},\,
\end{equation}%
where $t_{\mathbf{k}\sigma }=t_{\mathbf{k}}-\sigma IS$ is the mean-field
spectrum, $\Sigma _{\mathbf{k}\sigma }(E)$ is the self-energy. The latter
quantity can be expressed as an irreducible Green's function \cite{329}%
\begin{equation}
\Sigma _{\mathbf{k}}^{\sigma}(E)=\langle \!\langle \lbrack c_{\mathbf{k}\sigma},H_{%
\text{int}}]|[H_{\text{int}},c_{\mathbf{k}\sigma}^{\dagger }]\rangle \!\rangle
_{E}^{\text{irr}}  \label{irr}
\end{equation}%
where $H_{\text{int}}$ is the perturbation $s-f$  Hamiltonian with the
mean-field part being subtracted, `'irr\textquotedblright\ means that the
terms diverging as $(E-t_{\mathbf{k}\sigma })^{-n}$ should be omitted in the
following expansion.

To second order in $I$ we have
\begin{eqnarray}
\Sigma _{\mathbf{k}}^{\sigma }(E) &=&2I^{2}SR_{\mathbf{k}}^{\sigma },
\label{g1} \\
R_{\mathbf{k}}^{\uparrow }(E) &=&\sum_{\mathbf{q}}\frac{n_{\mathbf{%
k-q\downarrow }}^{{}}}{E-t_{\mathbf{k-q\downarrow }}+\omega _{\mathbf{q}}}%
,~R_{\mathbf{k}}^{\downarrow }(E)=\sum_{\mathbf{q}}\frac{1-n_{\mathbf{k-q}%
\uparrow }^{{}}}{E-t_{\mathbf{k-q\uparrow }}-\omega _{\mathbf{q}}}
\label{g2}
\end{eqnarray}%
with $n_{\mathbf{k}\sigma }^{{}}=f(t_{\mathbf{k}\sigma })$  the Fermi
function, $\omega _{\mathbf{q}}$ the magnon frequency (this is small as
compared to characteristic electron energies, but provides a cutoff for
logarithmic Kondo-like divergences).

Performing decoupling at the next step (which corresponds to summing up the
ladder diagrams \cite{RMP}) yields
\begin{equation}
\Sigma _{\mathbf{k}}^{\sigma }(E)=\frac{2I^{2}SR_{\mathbf{k}}^{\sigma }}{%
1+\sigma IR_{\mathbf{k}}^{\sigma }}  \label{ladd}
\end{equation}%
For $I>0$ we have in the half-metallic state (where $n_{\mathbf{k}\downarrow
}^{{}}=0$)

\begin{eqnarray}
G_{\mathbf{k\uparrow }}^{{}}(E) &=&G_{\mathbf{k\uparrow }}^{0}(E)=(E-t_{%
\mathbf{k}}+IS)^{-1}  \notag \\
G_{\mathbf{k\downarrow }}^{{}}(E) &=&G_{\mathbf{k\downarrow }}^{0}(E)=\left(
\ E-t_{\mathbf{k}}+IS-\frac{2IS}{1-IR_{\mathbf{k\uparrow }}(E)}\right) ^{-1}
\label{empty}
\end{eqnarray}%
Thus, the spin-up electrons move freely and spin-down states are incoherent
(the spectral density below the Fermi level  comes from branch cut of the
Green's function owing to the resolvent  $R_{\mathbf{k\uparrow }}(E)$, but
not from poles).

The result (\ref{empty}) is exact in the case of empty conduction band since
this corresponds to the exact solution of the two-body electron-magnon
scattering problem. In the limit $I\rightarrow +\infty $ Eq. (\ref{empty})
yields correct result (\ref{I+}) also for finite electron density. Thus for $%
I>0$ the approximation (\ref{ladd}) is quite satisfactory.

In the limit $I\rightarrow -\infty $ Eq.(\ref{empty}) gives correctly the
spectrum of spin-down quasiparticles (see Eq.(\ref{I+})). However, for \ $%
I<0~$the ladder approximation\ is not appropriate for describing NQP states,
although it reproduces a non-pole structure. There are several problems.
First, Eq.(\ref{ladd}) does not reproduce atomic limit and  limit of strong
correlations (see Appendix). Second, it contains the false Kondo pole, which
may lead to some problems.

Thus we use, instead of (\ref{irr}), another representation for the Green's
function, which is appropriate for describing NQP states. Writing down the
equations of motion we obtain%
\begin{eqnarray}
G_{\mathbf{k\uparrow }}(E) &=&G_{\mathbf{k\uparrow }}^{0}(E)+I^{2}[G_{%
\mathbf{k\uparrow }}^{0}(E)]^{2}\Phi _{\mathbf{k}}(E)  \label{nirr1} \\
\Phi _{\mathbf{k}}(E) &=&\sum_{\mathbf{p}}\Gamma _{\mathbf{kp}}^{{}}(E)
\notag
\end{eqnarray}%
with%
\begin{eqnarray}
\Gamma _{\mathbf{kp}}^{{}}(E) &=&\sum_{\mathbf{pr}}\langle \langle S_{%
\mathbf{p}}^{-}c_{\mathbf{k}-\mathbf{p\downarrow }}+\delta S_{\mathbf{p}%
}^{z}c_{\mathbf{k}-\mathbf{p\uparrow }}|c_{\mathbf{k+r\downarrow }}^{\dagger
}S_{\mathbf{r}}^{+}+c_{\mathbf{k}+\mathbf{r\uparrow }}^{\dagger }\delta S_{%
\mathbf{r}}^{z}\rangle \rangle _{E}  \notag \\
&=&\sum_{\mathbf{pr}}\langle \langle S_{\mathbf{p}}^{-}c_{\mathbf{k}-\mathbf{%
p\downarrow }}^{{}}|c_{\mathbf{k+r\downarrow }}^{\dagger }S_{\mathbf{r}%
}^{{}}+2c_{\mathbf{k}+\mathbf{r\uparrow }}^{\dagger }\delta S_{\mathbf{r}%
}^{z}\rangle \rangle _{E}  \label{nirr}
\end{eqnarray}%
Here $\delta A=A-\langle A\rangle$, we have used symmetry, purely
longitudinal fluctuation terms (without spin flips) being not important at
low temperatures. Note that the Green's function (\ref{nirr}) is not
irreducible. The first term in (\ref{nirr1}) does not give a contribution
to the density of states in the HMF state and, in any case, for large $|I|$.

The equation of motion for $\Gamma $ reads after decoupling%
\begin{eqnarray}
&&\left( E-t_{\mathbf{k-p}}-IS\right) \Gamma _{\mathbf{kp}}^{{}}(E)%
\begin{array}{c}
=%
\end{array}%
2Sn_{\mathbf{k}-\mathbf{p\downarrow }}-2\sum_{\mathbf{q}}\langle c_{\mathbf{%
k-q\uparrow }}^{\dagger }c_{\mathbf{k}-\mathbf{p\downarrow }}S_{\mathbf{p-q}%
}^{-}\rangle   \notag \\
&&-2ISn_{\mathbf{k}-\mathbf{p\downarrow }}\sum_{\mathbf{r}}\langle \!\langle
c_{\mathbf{k\uparrow }}|c_{\mathbf{k+r\downarrow }}^{\dagger }S_{\mathbf{r}%
}^{+}+2c_{\mathbf{k}+\mathbf{r\uparrow }}^{\dagger }\delta S_{\mathbf{r}%
}^{z}\rangle \rangle _{E}
\end{eqnarray}%
where $n_{\mathbf{k}\sigma }=\langle c_{\mathbf{k}\sigma }^{\dagger }c_{%
\mathbf{k}\sigma }^{{}}\rangle .$ Using the equation
\begin{equation}
(E-t_{\mathbf{k}}+IS)\sum_{\mathbf{r}}\langle \!\langle c_{\mathbf{k\uparrow
}}|c_{\mathbf{k+r\downarrow }}^{\dagger }S_{\mathbf{r}}^{+}+2c_{\mathbf{k}+%
\mathbf{r\uparrow }}^{\dagger }\delta S_{\mathbf{r}}^{z}\rangle \rangle
_{E}=-I\sum_{\mathbf{p}}\Gamma _{\mathbf{kp}}^{{}}(E)
\end{equation}%
we derive the integral equation%
\begin{equation}
\left( E-t_{\mathbf{k-p}}-IS\right) \Gamma _{\mathbf{kp}}^{{}}(E)=2S%
\widetilde{n}_{\mathbf{k}-\mathbf{p\downarrow }}(E)-I\frac{E-t_{\mathbf{k}%
}-IS}{E-t_{\mathbf{k}}+IS}\widetilde{n}_{\mathbf{k}-\mathbf{p\downarrow }%
}\sum_{\mathbf{q}}\Gamma _{\mathbf{kq}}^{{}}(E)  \label{int2}
\end{equation}%
with the renormalized occupation numbers

\begin{equation}
\widetilde{n}_{\mathbf{k\downarrow }}=n_{\mathbf{k\downarrow }}-\frac{1}{S}%
\sum_{\mathbf{q}}\langle c_{\mathbf{k-q\uparrow }}^{\dagger }c_{\mathbf{%
k\downarrow }}S_{\mathbf{q}}^{-}\rangle   \label{distr}
\end{equation}%
The second term in (\ref{distr}) is somewhat similar to that in the
paramagnetic Kondo problem (e.g., in the  Nagaoka decoupling  \cite{Nag}).
Calculating the corresponding Green's function%
\begin{equation}
\langle \!\langle S_{\mathbf{q}}^{+}c_{\mathbf{k-q\uparrow }}|c_{\mathbf{%
k\downarrow }}^{\dagger }\rangle \rangle _{E}=-2ISG_{\mathbf{k-q\uparrow }%
}^{0}(E)G_{\mathbf{k\downarrow }}^{0}(E)
\end{equation}%
and using the spectral representation we obtain%
\begin{equation}
\langle c_{\mathbf{k-q\uparrow }}^{\dagger }c_{\mathbf{k\downarrow }}S_{%
\mathbf{q}}^{-}\rangle =-2ISn_{\mathbf{k\downarrow }}/(E_{\mathbf{%
k\downarrow }}-t_{\mathbf{k-q}}+IS)
\end{equation}%
The spin-down distribution function can be calculated as%
\begin{equation*}
n_{\mathbf{k\downarrow }}=Z_{\mathbf{k\downarrow }}f(E_{\mathbf{k\downarrow }%
})
\end{equation*}%
where the residue of the Green's function $G_{\mathbf{k\downarrow }}^{0}(E)$ at the spin-polaron pole  $E_{\mathbf{k}\downarrow }$ is given by%
\begin{equation}
Z_{\mathbf{k\downarrow }}=\left( 1-\frac{\partial }{\partial E}\Sigma _{%
\mathbf{k}}^{\mathbf{\downarrow }}(E=E_{\mathbf{k\downarrow }})\right) ^{-1}
\end{equation}%
At large $|I|$ we can use the simple approximation (see Appendix)

\begin{equation}
Z_{\mathbf{k\downarrow }}=\frac{2S}{2S+1},~~\widetilde{n}_{\mathbf{%
k\downarrow }}=\left( \frac{2S+1}{2S}\right) ^{2}n_{\mathbf{k\downarrow }}=%
\frac{2S+1}{2S}f(E_{\mathbf{k\downarrow }})  \label{2S+1}
\end{equation}%
It should be noted that these calculations are not quite strict, but are in
fact in spirit of the quasiclassical $1/S$ expansion \cite{329}, treatment
of higher orders being needed for justification.

Eq.(\ref{int2}) can be easily solved. In  higher orders of perturbation theory
(taking into account next terms in the expansion) the denominators $\left(
E-t_{\mathbf{k-p}}-IS\right) ^{-1}$ are replaced by the exact Green's
functions $G_{\mathbf{k-p\downarrow }}^{0}(E)$, so that we derive%
\begin{eqnarray}
\Phi _{\mathbf{k}}(E) &=&\frac{2I^{2}S\widetilde{R}_{\mathbf{k}}^{\uparrow
}(E)}{1+I\widetilde{R}_{\mathbf{k}}^{\uparrow }(E)G_{\mathbf{k\uparrow }%
}^{0}(E)/G_{\mathbf{k\downarrow }}^{0}(E)}  \label{aa} \\
\widetilde{R}_{\mathbf{k}}^{\uparrow }(E) &=&\sum_{\mathbf{q}}\widetilde{n}_{%
\mathbf{k-q\downarrow }}(E)G_{\mathbf{k-q\downarrow }}^{0}(E+\omega _{%
\mathbf{q}})  \label{rrr}
\end{eqnarray}

This structure differs from the \textquotedblleft ladder\textquotedblright\
one (\ref{ladd}) by the presence of the factor $G_{\mathbf{k\uparrow }%
}^{0}(E)/G_{\mathbf{k\downarrow }}^{0}(E).$

\section{Results of numerical calculations}

The final result for the NQP contribution to the spin-up Green's function can be written
down in the form%
\begin{equation}
G_{\mathbf{k\uparrow }}(E)-G_{\mathbf{k\uparrow }}^{0}(E)=\frac{2ISG_{%
\mathbf{k\uparrow }}^{0}(E)}{[G_{\mathbf{k\downarrow }}^{0}(E)]^{-1}+[I%
\widetilde{R}_{\mathbf{k}}^{\uparrow }(E)G_{\mathbf{k\uparrow }}^{0}(E)]^{-1}%
}  \label{nqp}
\end{equation}%
Unlike (\ref{ladd}), expression (\ref{nqp}) has no false \textquotedblleft
Kondo\textquotedblright\ pole, so that analytical properties are fair.
Besides that, it yields  the correct structure in the large-$|I|$ limit (see
Appendix).

In our model, HMF state is characterized by large spin splitting which
exceeds the Fermi energy. Thus the usual band states are absent below the
Fermi level, and the NQP states  owing to branch cut of  the resolvent (\ref%
{rrr}) dominate in this region. The instability of HMF state is determined
by occurrence of  the pole of the Green's function (\ref{nqp}) below $E_{F}$, which  can take place with increasing the Fermi energy.

Despite the presence of the energy gap in the spin-up subband, the current
carriers are depolarized even in the ground state owing to the
electron-magnon interaction. Physically,  the occupied spin-down electron
states are a superposition of the states $\ |S\rangle |\mathbf{\downarrow }%
\rangle $ and $\ |S-1\rangle |\uparrow \rangle $, so that we can pick up
from these states spin--down and spin up electrons with the corresponding
spectral weights. The depolarization is maximum in the large-$|I|$ limit where (for $S=1/2$) it reaches 100\%.

\begin{figure}[tbp]
\includegraphics[width=3.3in, angle=0]{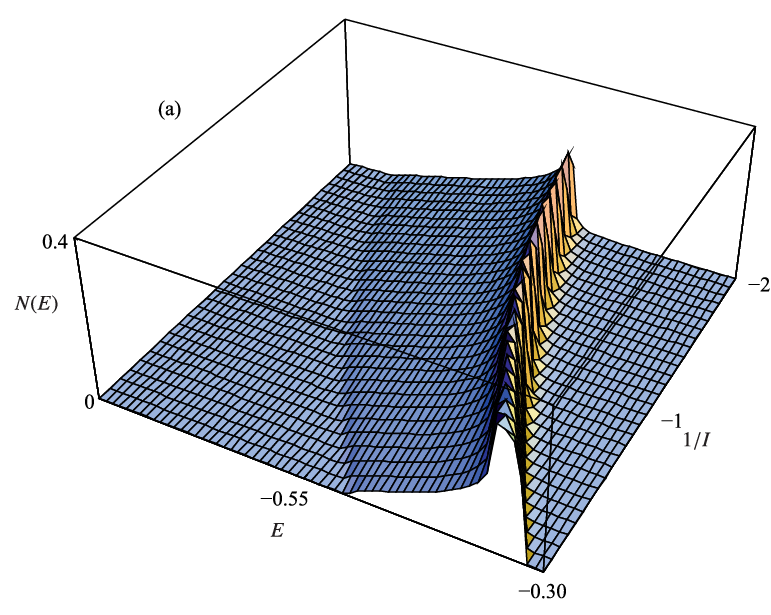}
\includegraphics[width=3.3in, angle=0]{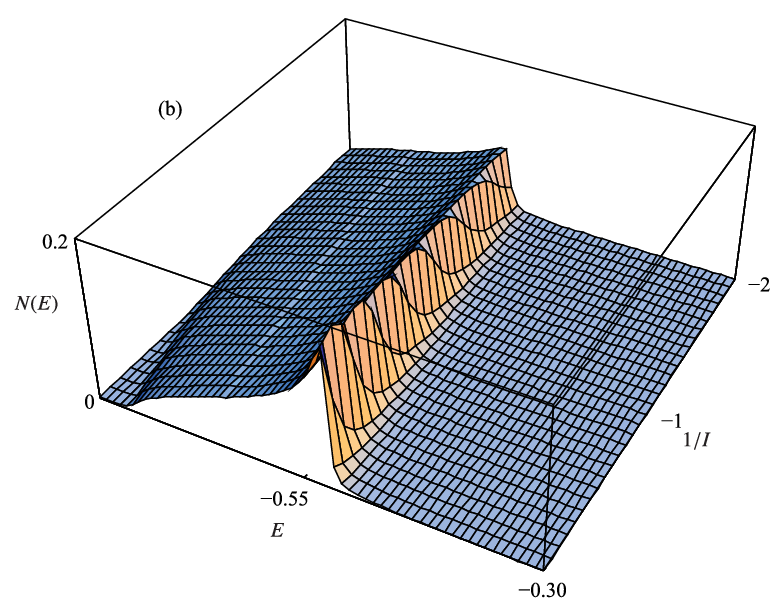}
\caption{Non-quasiparticle spin-up density of states $N(E)$ in a half-metallic ferromagnet for the semielliptic bare band with the width of $W=2$ and different $I$ values, $S=1/2$ (a) and $S=3/2$ (b) according to (\ref{nqp})}
\label{fig:1}
\end{figure}

\begin{figure}[tbp]
\includegraphics[width=3.3in, angle=0]{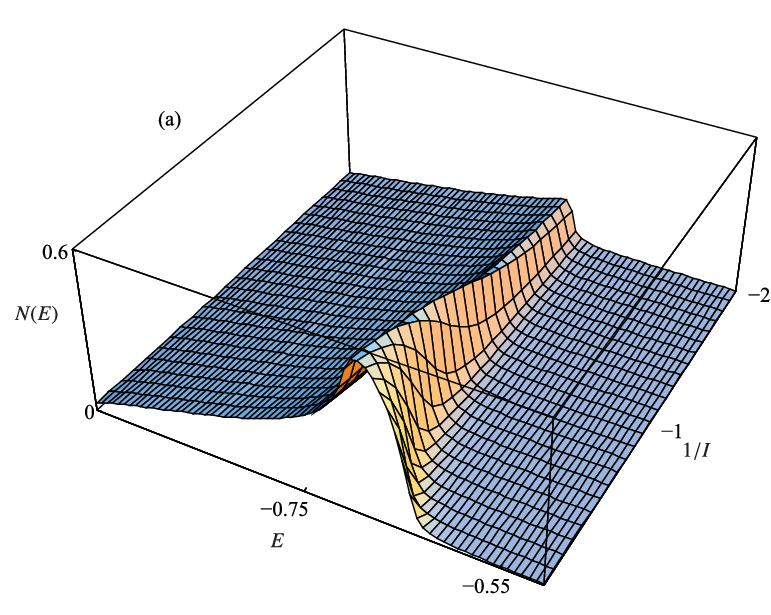}
\includegraphics[width=3.3in, angle=0]{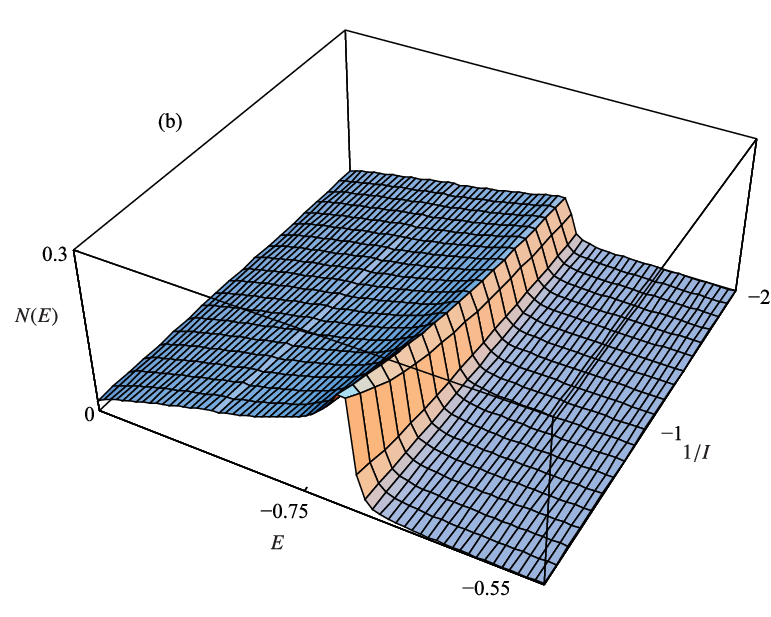}
\caption{Non-quasiparticle spin-up density of states in a half-metallic ferromagnet with different $I$ values $S=1/2$ (a) and $S=3/2$ (b) according to (\ref{ladd})}
\label{fig:2}
\end{figure}

Fig.1 shows the energy dependence of NQP density of states for different values of the $s-f$ exchange parameter, the occupation numbers being taken in the approximation (\ref{2S+1}).
The energy is calculated from the centre of the bare band shifted by atomic value $I(S+1)$.
The chemical potential corresponds to $-1< t_{\mathbf{k}}-I(S+1) < \mu_0 =-0.7$.
Note the shift of the true chemical potential $\mu$ with the change of $I$ owing to spin-polaron renormalization.
One can also see that the $|I|$ dependence of the NQP contribution size is weak since $|I|$ is in fact not a small parameter in the HMF regime (large spin splitting).


For comparison, the results according to Eq.(\ref{ladd}) are presented in Fig.2.
The chemical potential corresponds to $-1< t_{\mathbf{k}}-IS < \mu_0 =-0.7$.
One can see that the visual picture is quite different, especially in the quantum (small $S$) case.
In particular, this approximation does not yield band narrowing (which is $2S/(2S+1)$ in the large-$I$ case). A complicated behavior at intermediate $I$ for $S=1/2$ demonstrates the influence of the Kondo singularity.

Note that in fact the Kondo divergence has no crucial effect because of the one-side character of the singularity, so that $N_\uparrow(E_F)=0$; besides that, smearing due to spin dynamics should be taken into account. Thus a dependence on cutoff at the spin-wave frequency arises, which is stronger in the approximation (\ref{ladd}).

\section{Discussion and conclusions}

The problem of half-metallic ferromagnetism in the $s-f$ exchange model is somewhat similar to the Kondo lattice problem in paramagnetic (or weakly magnetic) state, but can be investigated rigorously by rather simple method of the Green's functions.

In the present work, we have obtained a consistent solution of the Kondo problem in the case of strong (half-metallic) ferromagnetism in terms of NQP states, the ultimately strong correlation limit being correctly described. In our case, there are no Kondo corrections to the ground state magnetic moment, so that ferromagnetism remains saturated.
On the other hand, in the case of weak magnetism the standard scaling theory starting from the approximation (\ref{ladd}) describes successfully the competition of the Kondo effect (screening of magnetic moments) and intersite exchange interactions \cite{IK}. Note, however, that half-metallic solutions with a Kondo-reduced magnetic moment can occur in the strong-coupling regime \cite{IK91}.

The Kondo divergences in a ferromagnetic state were also considered for the usual magnetic \cite{IK02} and pseudospin \cite{Trefilov} problems, transport properties (e.g., sharp energy dependence of resistivity and large thermoelectric power in the case of impurity scattering) being discussed.

The HMF spectrum picture obtained includes to first line non-quasiparticle states which are the only electron states near the Fermi level for one of spin projections. The example of HMF turns out here instructive to investigate a very important problem of passing from one-electron to many-electron statistics.

The NQP states were classified as minority ones (above the energy gap, e.g., in NiMnSb and CrO$_2$ \cite{NQP}) and majority ones (below the energy gap, e.g., in Mn$_2$VAl \cite{NQP1}). The latter situation just corresponds to the case $I<0$.
Thus our treatment can be used in a combined consideration with first-principle band calculations.

The author is grateful to M. I. Katsnelson and A. I. Lichtenstein for useful discussions.
This work was supported in part by the Division of Physical Sciences and Ural Branch of Russian Academy of Sciences
(project no. 15-8-2-9).

\section*{Appendix. Atomic and strong correlation limits}

The atomic-limit Green's functions can be calculated exactly for arbitrary
electron density $n$ and magnetization \cite{RMP}. They have a two-pole
structure corresponding to two atomic levels%
\begin{equation}
G_{{}}^{\sigma }(E)=\frac{1}{2S+1}\left( \frac{S+n+\sigma \langle S\rangle }{%
E+IS}+\frac{S+1-\sigma \langle S\rangle -n}{E-I(S+1)}\right)
 \label{at}
\end{equation}%
Note that a part of spectral weight vanishes in the large-$|I|$ limit.

The strong correlation limit $|I|\rightarrow \infty $ can be also
investigated rigorously \cite{orb}. The solution  can be found by using the
atomic representation of Hubbard's X-operators. (However, for finite $|I|$
such analytical calculations in a closed form are difficult and not
instructive: there are too many states with different spin projections and
occupations.)

In the limit $I\rightarrow +\infty $ we have%
\begin{eqnarray}
G_{\mathbf{k}}^{\uparrow }\left( E\right)  &=&\frac{1}{\epsilon -t_{\mathbf{k%
}}},~G_{\mathbf{k}}^{\downarrow }\left( E\right) =\left[ \epsilon -t_{%
\mathbf{k}}+\frac{2S}{R_{\mathbf{k}}(\epsilon )}\right] ^{-1}  \label{I+} \\
R_{\mathbf{k}}(\epsilon ) &=&\sum_{\mathbf{q}}\frac{1-f(t_{\mathbf{k-q}%
}^{{}})}{\epsilon -t_{\mathbf{k-q}}^{{}}-\omega _{\mathbf{q}}}
\end{eqnarray}%
with $\epsilon =E+IS$.

In the limit $I\rightarrow -\infty $ we derive
\begin{equation}
G_{\mathbf{k}}^{\downarrow }\left( E\right) =\frac{2S}{2S+1}(\epsilon -t_{%
\mathbf{k}}^{\ast })^{-1}  \label{I-}
\end{equation}%
\begin{eqnarray}
G_{\mathbf{k}}^{\uparrow }\left( E\right)  &=&\frac{2S}{2S+1}\left[ \epsilon
-t_{\mathbf{k}}^{\ast }+\frac{2S}{R_{\mathbf{k}}^{\ast }(\epsilon )}\right]
^{-1},  \label{nqpl} \\
R_{\mathbf{k}}^{\ast }(\epsilon ) &=&\sum_{\mathbf{q}}\frac{f(t_{\mathbf{k-q}%
}^{\ast })}{\epsilon -t_{\mathbf{k-q}}^{\ast }+\omega _{\mathbf{q}}}
\end{eqnarray}%
with $\epsilon =E-I(S+1),t_{\mathbf{k}}^{\ast }=[2S/(2S+1)]t_{\mathbf{k}}$.

The Green's function (\ref{nqpl}) has no poles, at least for small current
carrier concentration, and the whole spectral weight of minority states is
provided by the branch cut (non-quasiparticle states). The one-electron
distribution functions read

\begin{equation}
n_{\mathbf{k\downarrow }}=\frac{2S}{2S+1}f(t_{\mathbf{k}}^{\ast }),~n_{%
\mathbf{k}\uparrow }\simeq \frac{n}{2S+1},
\end{equation}%
so that $ n_{\mathbf{k}\uparrow }$ is very weakly ${\bf k}$-dependent.
One can see that the equation for the chemical potential, $\sum_{\mathbf{k}%
\sigma }n_{\mathbf{k}\sigma }=n$, is satisfied just owing to NQP spin-up states.

It should be noted that  the  motion of  observable strongly correlated quasiparticles
are  described by many-electron projection X-operators,%
\begin{equation*}
c_{i}^{\dagger }\rightarrow \sqrt{\frac{2S}{2S+1}}x_{i}^{\dagger },\
x_{i}^{\dagger }=|i,S-1/2,S-1/2\rangle \langle i,S,S|
\end{equation*}
where $\ |i,S^{\prime },M\rangle $ is the state on a site $i$ with total
spin $S^{\prime} $ and its projection $M$. The corresponding distribution
function   does not contain the factor of $2S/(2S+1)$.

\begin{equation}
\langle \!\langle x_{\mathbf{k}}|x_{\mathbf{k}}^{\dagger }\rangle \!\rangle
_{E}=1/(\epsilon -t_{\mathbf{k}}^{\ast }),~\langle x_{\mathbf{k}}^{\dagger
}x_{\mathbf{k}}^{{}}\rangle =f(t_{\mathbf{k}}^{\ast }).
\end{equation}



\end{document}